\documentstyle[prl,aps,multicol]{revtex}
\begin{document}
\preprint{TNT 95-GlCM-V.5}
\draft

\title{Broken ergodicity and glassy behavior \\ in a deterministic
       chaotic map}

\author{A. Crisanti, M. Falcioni\cite{INFN}
        and A. Vulpiani\cite{INFN}}

\address{Dipartimento di Fisica, Universit\`a ``La Sapienza'',
         P.le A. Moro 2, 00185, Roma, Italy}

\date{\today}

\maketitle

\begin{abstract}
A network of $N$ elements is studied in terms of a deterministic 
globally coupled map which can be chaotic.
There exists a range of values for the parameters of the map where
the number of different macroscopic configurations ${\cal N}(N)$ 
is very large, ${\cal N}(N) \sim \exp \, \sqrt{c(a) N}$
and there is violation of selfaveraging.
The time averages of functions, which depend on a single 
element, computed over a time $T$, have probability distributions 
that do not collapse to a delta function, for increasing $T$ 
and $N$. This happens for both chaotic and regular motion, 
{\it i.e.} positive or negative Lyapunov exponent.
\end{abstract}

\pacs{05.45.+b,
      05.90.+m}

\begin{multicols}{2}
Glassy behavior is still one of the central subjects of modern 
physics \cite{sping}. Among the most interesting 
features are the violation of
time traslation invariance, and fluctuation-dissipation theorem, 
even for very long times \cite{flure}. This follows from the breaking of
ergodicity in the phase space caused by a rough energy 
landscape. In spin glasses, the paradigmatic system, the complex
dynamics follows from frustration and randomness. It is, however, 
obvious that these are not a prerequisite for glassy behavior,
think for example to ordinary window glass. Recently an 
increasing amount of work has been devoted to the study of
glassy behavior in non random systems \cite{nonran}. In this 
case the complex behavior follows from the existence of a 
large number of metastable states which slows 
down the dynamics. 

We shall concentrate on systems of globally coupled elements
without any intrinsic disorder. 
Dynamical systems of this type have been 
studied recently, in great details, in many fields \cite{Kanboo},
and used to model a wide class of 
phenomena, such as Josephson-junction array, evolution dynamic 
of ecosystems, neural-networks for neurodynamics.

The interesting point is that the macroscopic description of these
systems may be extremely rich. The variety of the macroscopic 
description can be taken as an indication of complexity \cite{Parisi}.

In this Letter we shall study a network of elements -- a 
globally coupled map -- given by:
\begin{equation}
\label{eq:uno}
x_i(n+1)=(1-\epsilon)f(x_i(n))+{\epsilon \over N} \sum_{j=1}^N f(x_j(n))
\end{equation}
where $n$ is the discrete time variable and $i=1,\cdots,N$ is 
the index identifying the elements. For $f(x)$ we choose the logistic 
map 
\begin{equation}
\label{eq:due}
f(x)=1-a\, x^2.
\end{equation}
The system (\ref{eq:uno}), that has been thoroughly studied 
by Kaneko \cite{Kan}, displays a very rich behavior. Interestingly, 
for certain values of the parameters $a$ and $\epsilon$, one 
observes a violation of the law of large numbers: a subtle 
coherence among the elements $x_i$ induces anomalous 
fluctuations of the mean field 
\begin{equation}
\label{eq:tre}
h(n)= {1\over N} \sum_{j=1}^N f(x_j(n)), 
\end{equation}
so that the mean square deviation, 
$\langle h^2 \rangle - \langle h \rangle ^2 ,$ decays as 
$N^{-\beta}$, with $\beta < 1 .$
Moreover, the elements split into clusters whose probability 
distributions resembles, for suitable values of the parameter, 
to those derived from the theory of spin glasses \cite{Kan91}. 

In this letter we show that the analogy of this system with 
spin glasses, previously proposed by Kaneko, is very deep: 
for a certain range of values of the control 
parameter $a$, the system exhibits a very large numbers of 
``macroscopic" states and a violation of selfaveraging, 
indicating that the system is ``complex'' in the above
stated sense.  

Because of the global coupling in 
(\ref{eq:uno}), one can have the phenomenon of clusterization: 
if, at a certain time $\widetilde{n}$, the variables of some 
elements take on identical values, they will maintain equal 
values for every $n > \widetilde{n}$. Therefore, it is possible 
to introduce a ``macroscopic" description of the system in terms of the 
number $N_i$ of elements in the $i^{\rm th}$ cluster and to identify a 
macroscopic state by the set of integers 
$\left\{ N_i \right\} = 
\left[ N_1,N_2,\dots,N_m | \sum_{i=1}^m N_i =N \right]$, 
that characterize the population of each cluster. All the 
microscopic states with the same cluster structure, i.e. with 
identical values for the $\left\{ N_i \right\}$, but 
differing either by a permutation of some $x_j$ or by the 
value of the variables $x_j$ in a particular cluster, belong 
to the same macroscopic state (by definition). 

We start by estimating the number of 
relevant macroscopic states of the system as a function of 
$N$, for different values of $a$. This is done by generating 
for each value of $N$ and $a$ a set ${\cal M} \gg 1$ of initial 
conditions $x_i^{(\alpha)}(0),$ with $\alpha =1,2,\cdots,{\cal M}$, 
selecting at random each component $x_i^{(\alpha)}(0)$ 
with uniform distribution in the interval $[-1,+1]$;
we used ${\cal M} = 10^4 \div 5\times 10^4$. 
The numerical results reported here were obtained with $\epsilon=0.1$.
For each initial condition the map (\ref{eq:uno}) is
iterated and the final ``cluster structure'' recorded.
In all cases we have discarded $6\cdot 10^4$ thermalization 
steps before performing measurements. From each set of initial 
conditions we obtained a "cluster structure", i.e. a set 
$\left\{ N_j^{(\alpha)}  \right\},$  and, from the frequency 
of occurrence of every set, we estimated the probability, 
$P\left( N_1,N_2,\dots,N_m \right),$ to observe the macroscopic 
state $\left( N_1,N_2,\dots,N_m \right)$. 
To take into 
account only the relevant states, {\it i.e.} to neglect the 
configurations with very small probability, we compute the entropy
\begin{eqnarray}
\label{eq:qua}
 S(N)= - \sum_{\left( N_1,N_2,\dots,N_m \right)} 
 &P_N& (N_1,N_2,\dots,N_m ) \times \nonumber \\ \ln  &P_N& (N_1,N_2,\dots,N_m)
\end{eqnarray}
from which we extract the number of macroscopic configurations with 
non negligible probability
\begin{equation}
\label{eq:cin}
{\cal N}(N) = e ^{S(N)}. 
\end{equation}
A clusterization of $N$ elements in the language of number theory
is a partition of the integer $N$, i.e., a possible way to obtain $N$ by 
summing integers -- not necessarily all different. An upper bound 
for $S(N)$ and a guess for how it scales with the system size 
can be obtained noting that the asymptotic estimate for the number 
of partitions of $N$, ${\cal N}_p(N)$, is \cite{Abra}
\begin{equation}
\label{eq:sei}
{\cal N}_p(N) \simeq { e ^{\pi \sqrt{2N/3}} \over 4 N \sqrt{3} }. 
\end{equation}
From this we may guess that 
\begin{equation}
\label{eq:set}
S(N)^2 \simeq c(a) N +cost 
\end{equation}
with $c(a) \le c_{max}=(2/3)\pi^2 \simeq 6.58$. 
The constant $c(a)$ may be thought 
as an indicator of the fraction of the ``active'' or ``relevant''
elements in the system. In fact, if one defines the number 
$N_{act}$ by the relation 
\begin{equation}
\label{eq:ott}
{\cal N}(N) \sim e ^{\pi \sqrt{2N_{act}/3}},  
\end{equation}
it readily follows that
\begin{equation}
\label{eq:nov}
{c(a) \over c_{max}} = { N_{act} \over N}.  
\end{equation}

In Fig. \ref{fig:fig1} we display $S(N)^2$ as a function of $N$ 
for different values of $a$. For each initial condition we 
checked that the cluster structure, found at the end of the 
thermalization, is a ``permanent" structure, that was exactly 
the same after $6\times 10^4$ and $1.2\times 10^5$ successive 
iterations.

Figure \ref{fig:fig2} shows $c$ and  
the mean value $\langle \lambda \rangle$ 
of the maximum Lyapunov exponent of the system
averaged over the initial conditions. The bars on 
$\langle \lambda \rangle$  indicate the variance 
$\sigma_N$. For $a > 1.63$ the variance $\sigma _N$ 
goes to zero as $N$ increases and the probability 
distribution of the Lyapunov exponents, $P_N (\lambda)$, 
approaches a delta 
function, around some $\lambda > 0$, as $N \to \infty$. 
On the other hand, for $a < 1.63$  and for $N\to \infty$  
$\sigma _N$ approaches a non zero value, while
$P_N (\lambda)$ approaches a nontrivial asymptotic distribution.  

It is evident that for $a$ in 
the range $1.5 \div 1.7$, the system has a very large 
number of macroscopic states, while 
outside this range one detects only very few macroscopic states. 
We stress that we observe many macroscopic states -- both chaotic and regular --
not only many microscopic states, as in Ref. 
\cite{Wiese} where, in a system of $N$ globally coupled nonlinear 
oscillators, $(N-1)!$ stable limit cycles are found.  
Moreover, the maximum of $c(a)$ is attained in the region
where $\langle \lambda \rangle$ changes sign. 

The above results and the observation of the histograms of the 
maximum Lyapunov exponent, suggest the existence of four 
distinct regions in the behavior of the system. These are 
roughly equivalent to those identified by Kaneko. We stress, 
however, that whereas Kaneko's classification is based on 
microscopic considerations, ours follows from the observation 
of a macroscopic behavior. The four regions are: 
\begin{enumerate}
\item  
          $a < 1.5$, standard regular 
          motion, with $\langle \lambda \rangle < 0$ and few 
          macroscopic states.

\item 
          $1.5 < a < 1.63$,  ``glassy''
          regular motion, {\it i.e.}: $\langle \lambda \rangle < 0$ 
          and many macroscopic states. 

\item
          $1.63 < a < 1.70$, ``glassy'' 
          chaotic motion, with $\langle \lambda \rangle > 0$ and 
          many macroscopic states. 

\item
         $ a > 1.70$, standard chaotic 
         motion, with $\langle \lambda \rangle > 0$ and few 
         macroscopic states. 
\end{enumerate}
The analogy with the glassy behavior in disordered systems is not 
limited to the number of macroscopic states: in fact, in its 
glassy phase, system (\ref{eq:uno}) is characterized by a violation of the 
selfaveraging. Given an initial condition, ${\bf x}(0)$, we may compute 
the time average, over a time $T$, of the variable $x_i$: 
\begin{equation}
\label{eq:die}
\langle x_i\rangle_T = {1 \over T} \sum_{n=1}^T x_i (n)
\end{equation}
or of any other function of a single element.
In a standard situation, the probability distribution of 
$\langle x_i\rangle_T$ is expected to collapse to a delta function 
as $T$ and $N$ increase, that is, we expect that the time average 
over a very long time record of one element chosen at random inside a large 
system gives a complete description of the one-variable 
properties of the system. 

Figure 3a shows the probability distribution of $\langle x_i\rangle_T$, 
obtained by gathering the average values for different $i$ and 
different initial conditions, in the phase with standard chaos. 
By increasing the observation time $T$, the variance decreases and 
one recover the usual behavior. 

Figure 3b shows the situation in the case of the chaotic glassy
phase: the increasing of $T$ does not modify the nontrivial 
structure of the distribution. A similar behavior is observed
in the regular glassy phase.

These results, which are size independent, show that, 
in the glassy phase, the ergodicity is broken and the 
selfaveraging property does not hold, in other words: the 
time average on a single element is not sufficient for  
a complete characterization of the one-element statistics. 

We conclude with a brief discussion about the role of noise. 
Of course, for finite values of $N$, the presence of noise will 
restore ergodicity after a very long time, which depends on $N$. 
Therefore it is interesting to consider the behavior of the 
system subject to noise on large but finite times. 
In Ref. \cite{Wiese} for large $N$ even a small amount of 
noise is sufficient to induce transitions among the different 
attractors. On the contrary we present a different picture. 
Within our approach, the scenario described above for the 
system (\ref{eq:uno}) is very robust with respect to noise. 
In fact, the inclusion of a noise term $\sigma\eta_i(n)$ -- where
$\eta_i$ are independent random variables uniformly distributed 
in the interval $[-1,1]$ -- to the r.h.s. of equation (\ref{eq:uno}) 
does not change the probability distributions of $\langle x_i\rangle_T$ 
showed in Fig. 3a for noises strength as large as 
$\sigma \approx 10^{-2}$. Therefore the macroscopic states are stable 
for long times. This feature closely resemble to the behavior 
observed in glassy systems below the glassy transition temperature.

In conclusion we have shown that even simple models which do not
contain any intrinsic disorder can have an extremely rich behavior
at a macroscopic level. A behavior typical, e.g., of glasses.
This scenario is very robust with respect to small external noise, 
indicating a ``genuine'' property of the system.

\acknowledgments
We thank U. Marini Bettolo Marconi for useful discussions and for
critical reading of the manuscript.

\narrowtext

\begin{figure}
\caption{Squared entropy $S^2$ as a function 
         of the number of globally interacting elements $N$, for
         values of the parameter $a$ equal to $1.50$, $1.55$ and $1.63$.
         The lines are the linear best fits.
        }
\label{fig:fig1}
\end{figure}

\begin{figure}
\caption{Entropy coefficient $c$ (boxes) and average Lyapunov exponent 
         $\langle \lambda \rangle$ (triangles) as functions of $a$.
         The values of $\langle \lambda \rangle$ and of $\sigma_N$ 
         are computed with $N$ ranging from $100$ up to $3200$, 
         and with ${\cal M}=10^4$ initial conditions.
        }
\label{fig:fig2}
\end{figure}

\begin{figure}
\caption{Probability distribution of $\langle x_i\rangle_T$,
         at various observation times $T$ in the region of 
         standard chaos -- $a=1.80$ -- (a) and in the region 
         of glassy chaos -- $a=1.67$ -- (b). The different curves 
         correspond to averages performed, after a transient of 
         $6\times 10^4$ steps, on time intervals $T=0.75\times 10^4$,
         $1.5\times 10^4$, $3\times 10^4$, $6\times 10^4$, $12\times 10^4$. 
         For each curve ${\cal M}=10^3$ initial conditions 
         were considered with $N=300$.  The histograms do not change 
         qualitatively by changing $N$ up to $1200$.
        }
\label{fig:fig3}
\end{figure}

\end{multicols}
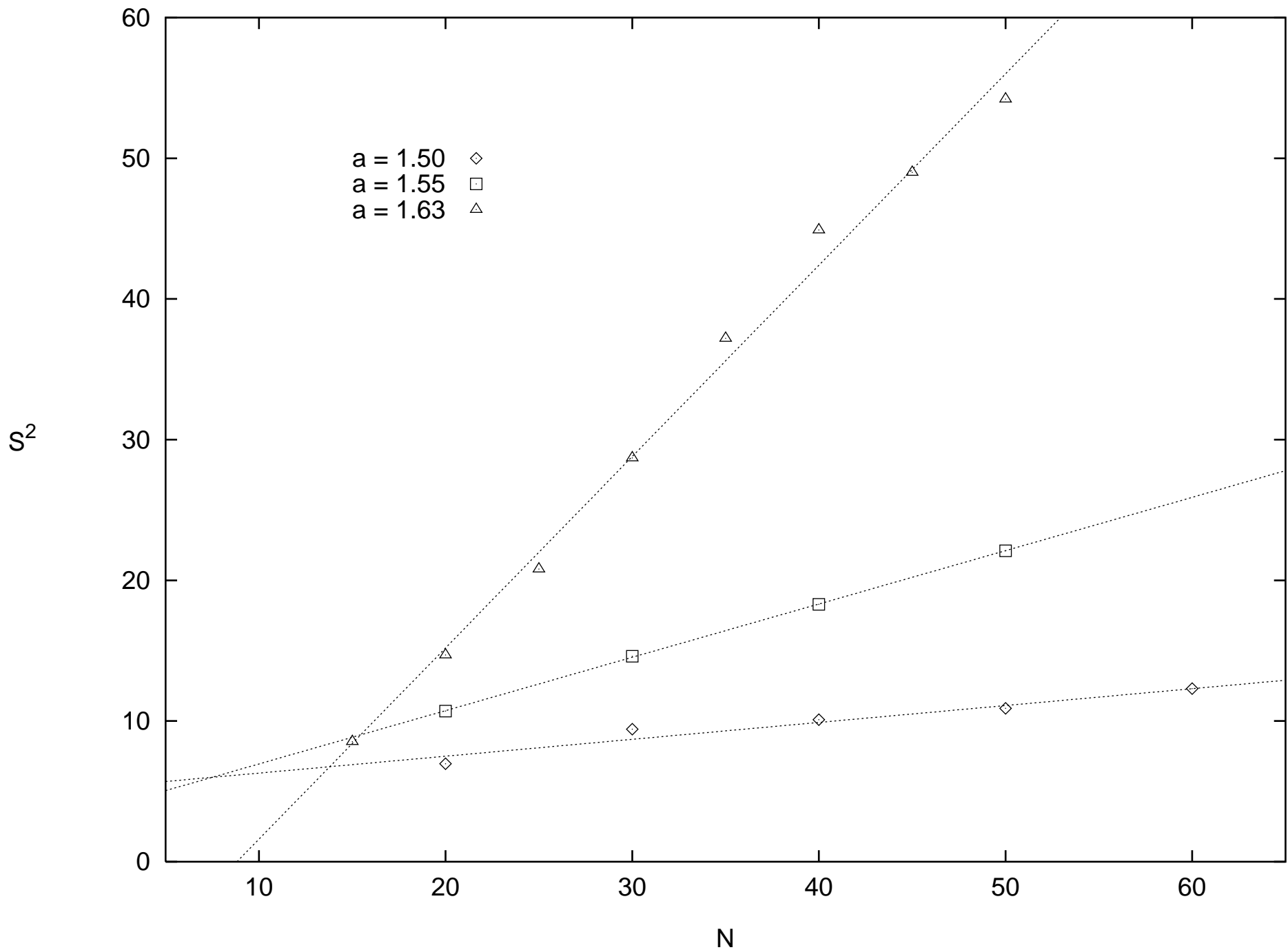
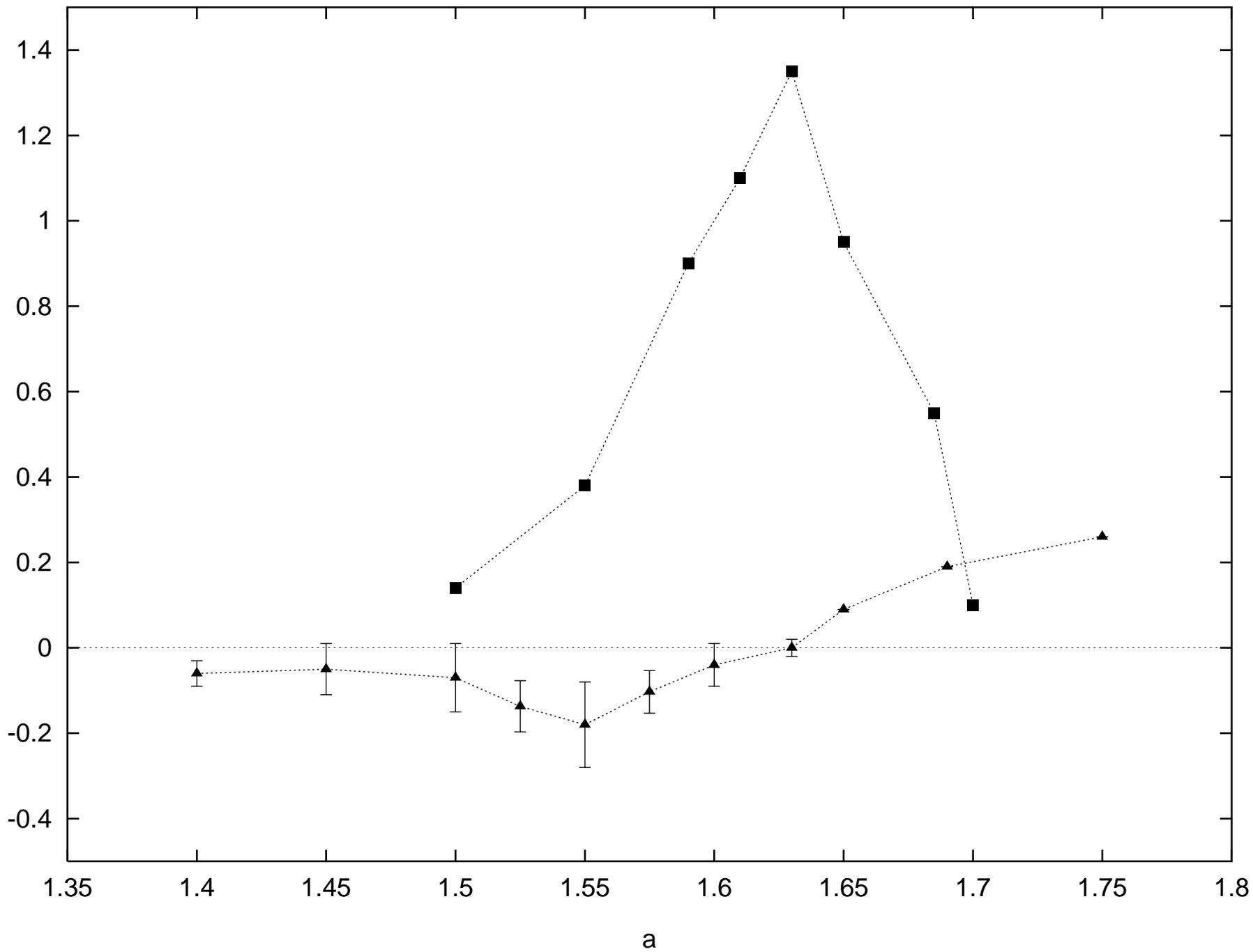
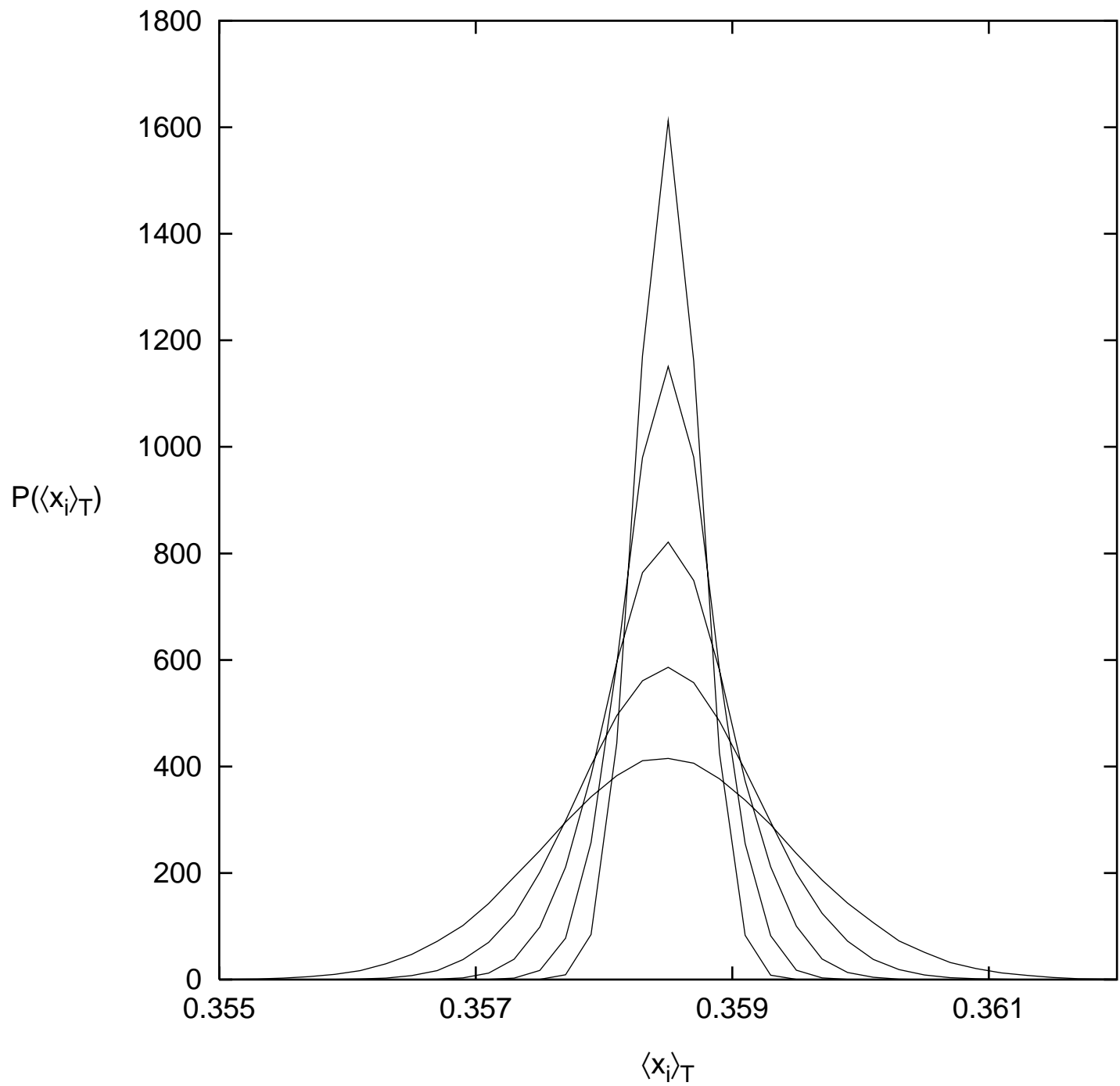
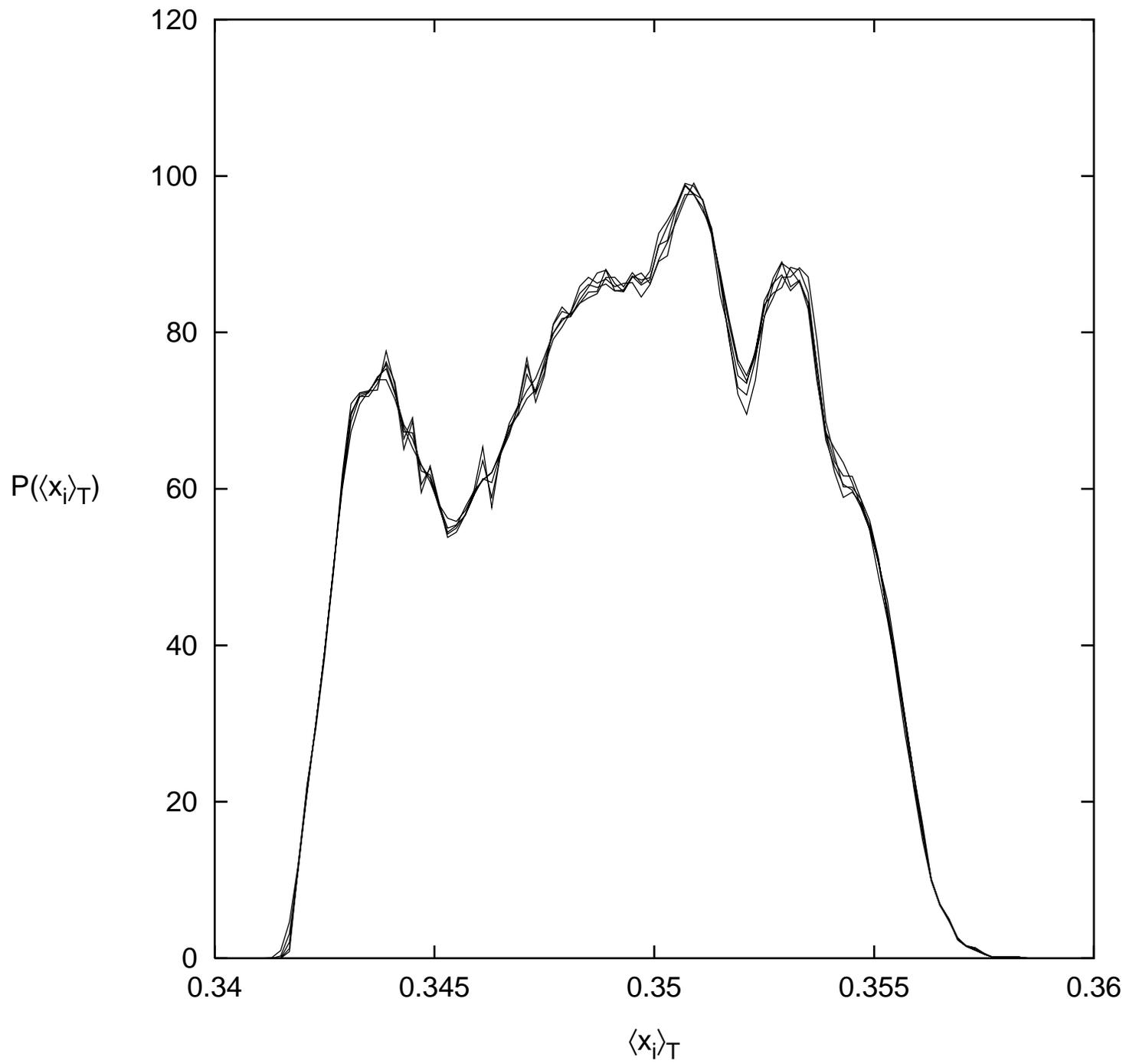
\end{document}